\documentclass[12pt,a4]{article}
\usepackage{amsmath,amssymb}
\usepackage{graphics,epsfig}
\Large
\begin{document}

\title{\bf{Perturbative approach to the hydrogen atom in strong magnetic field}}
\author{V. A. Gani, A. E. Kudryavtsev, V. A. Lensky, V. M. Weinberg}
\date{ }
\maketitle

\begin{abstract}
  The states of hydrogen atom with principal quantum number $n\leq 3$ and zero magnetic quantum number in constant 
homogeneous magnetic field $\mathcal H$ are considered. The perturbation theory series is summed with the help of Borel 
transformation and conformal mapping of the Borel variable. Convergence of approximate energy eigenvalues and their 
agreement with corresponding existing results are observed for external fields up to $n^3\mathcal H \sim 5$.
The possibility of restoring the asymptotic behaviour of energy levels using perturbation theory coefficients is also discussed.
\end{abstract}

\vspace{1.0cm}
The own magnetic fields of some astrophysical objects reach very high values~\cite{Alpar, Fass}. If we are interesting of 
the atomic spectra in these external  fields, it is convenient to introduce natural measure of field strength -- the atomic magnetic
field $\mathcal H_0\equiv e^3m^2c/\hbar^3=2.55\times 10^9$ G. 
The fields $\mathcal H$ up to one half of $\mathcal H_0$
are detected in vicinity of some white dwarfs.
Neutron stars possess
fields up to $\sim 10^4\mathcal H_0$. For correct interpretation of the observations results it is desirable to know the 
atomic hydrogen spectrum in this range of external fields. For this aim, computations based on adiabatic approach
with Landau level as initial approximation
were accomplished~\cite{Potekhin}. 
It will be shown here which 
part of the desired external field range could be covered with the help of the usual expansion in powers of $\mathcal H$,
starting from the Coulomb levels of hydrogen atom. 
We involved in the computations many orders of perturbation theory (up to 75th order). Summation of the series was performed with the 
help of Borel transformation, supplemented by conformal mapping of Borel variable.

    The Borel summation method was introduced into quantum field theory long enough (see e.g.~\cite{Z.-J.}). It has been 
tested on some quantum-mechanical problems (one of many examples is described in~\cite{Pop.Wein.}) and continues to find 
applications in modern works~\cite{Suslov}. Large hopes on the possibility to advance into strong coupling region were 
related with Borel summation of the perturbation series. Some rather simple problems, in which details can be traced and 
compared with corresponding exact results, supported this optimism. For example, for the funnel potential, $V(r) = -1/r + gr$, 
by applying conformal mapping of the Borel variable and Pad\' e-summation of the Borel transformant, the ground state energy 
at $g\rightarrow \infty $ was obtained in the form $E(g) = Cg^\nu $ with $\sim 0.2$ \% precision for index $\nu $ and $\sim 5$ \%
precision for coefficient $C$~\cite{Pop.Wein.}. It became clear later that such a successful summation presents a special but 
not the general case. One can guess that this success is a consequence of simplicity of this problem.
In contrast, asymptotic behaviour of energy levels in Stark and Zeeman effects comes into action at very large 
external fields values. For the Stark effect it is practically impossible now to reach the region of true asymptotic by 
perturbation series summation. An intermediate linear asymptotic is observed instead~\cite{Kaz.Pop., Wein.M.P.S.}. 

To introduce notations and scale we write down the Hamiltonian 
\begin{equation}
H = - \frac{1}{2}\nabla ^2 - \frac{1}{r} + \frac{1}{8}g(r^2-z^2)\equiv \hat H_0 + g\hat H_1.
\end{equation}
Here $g\equiv \mathcal H^2$, and hereafter we use units $\hbar=c=m=e=1$. 
In (1) we drop the elementary contribution of electron's spin and consider only states with magnetic quantum
number $m\equiv 0$. We can expand $E(g)$ as a formal series in powers of $g$:
\begin{equation}
E(g) = \sum_{k=0}^{\infty }E_k g^k .
\end{equation}
Now, we have to obtain hypersucceptibilities $E_k$. We could use the moment method for this aim.
This method is especially useful in the cases when variables 
in the Schr\" odinger equation can not be separated. Obviously the Zeeman effect presents just such a problem.
In the previous work~\cite{WGK} the moment method was applied to recurrent evaluation of hypersucceptibilities. 
Somewhat different version of the moment method was introduced in the work~\cite{Fernandez}.
 
For the four lower "isolated" hydrogen levels we immediately use here the results of ref.~\cite{WGK}. Unfortunately the 
computer code, employed in the work~\cite{WGK} for the relatively more complicated case of degenerate $3s$ and $3d$ states, contained
a mistake\footnote{We are thankful to Prof. V. D. Ovsyannikov for drawing our attention to this mistake.}.
Therefore we carried out new computation of $3s$ and $3d$ hypersucceptibilities. Results of computation for some orders are presented in
Table 1. These results are in agreement with results of the work~\cite{Silverstone} where high-order hypersucceptibilities were obtained at
the first time (but the method used in~\cite{Silverstone} is much more complicated than the moment method is).\\

\begin{center}
{\bf Table 1}
\vspace{0.5cm}

\begin{tabular}{|| l |l ||}
\hline
\multicolumn{2}{||c||}{\bf{Hypersucceptibilities of degenerate states}}\\
\hline
$k$ \medskip & $E_k$\,\it{ for 3s state}\\ 
\hline
           1 &  19.57851476711195477229924488394\\      
           2 & -7992.558488642566993349104381687\\      
           3 &  9951240.466276842310264046307800\\      
           4 & -20931559882.53444368634980579917\\      
           5 &  58826900682409.79349115290157121\\      
         25 &  1.3793233851820609414463787913215$\times 10^{94}$\\
         50 & -9.3227132696889616617788676903516$\times 10^{211}$\\
         75 &  2.8053533970811704326574930831176$\times 10^{340}$\\
\hline
$k$ \medskip & $E_k$\, \it{for 3d state}\\
\hline
           1 &  5.171485232888045227700755116050\\      
           2 & -1017.425886357433006650895618312\\      
           3 &  738127.8247387826897359536921995\\      
           4 & -923576528.5544112941189442008231\\      
           5 &  1677908319019.727217770438272530\\      
         25 &  1.0431217771758614011812311858395$\times 10^{92}$\\
         50 & -6.0721978561446884300072726553011$\times 10^{209}$\\
         75 &  1.7302552995055432680731087635037$\times 10^{338}$\\
\hline
\end{tabular}
\vspace{0.5cm}\\
\end{center}

As the order $k$ increases, hypersucceptibilities grow as a factorial~\cite{Avron}
\begin{equation}
E_k \rightarrow \tilde{E}_k = (-1)^{k+1}C_{nl}a_n^k\Gamma (2k + \beta _{nl}) ,
\end{equation}
where $a_n = (n^2/\pi )^2$,  $\beta_{nl}=2n-1+\frac{(-1)^l}{2}$, and $C_{nl}$ are not essential for us; one can find values of them in~\cite{WGK}
and references therein.
Eq.~(3) implies that series (2) is asymptotical and the formal sum of such 
a series is ambiguous. But in fact the choice of the summation method is restricted: physical considerations impose 
analytical properties of the function $E(g)$, which the true sum of series (2) is to reproduce. In the 
unphysical region, at $g < 0$, the diamagnetic perturbation $g\hat H_1$ changes its sign, the total 
Hamiltonian becomes "open" and the possibility of a spontaneous ionization of the atom appears. Therefore energy
eigenvalue should have imaginary part at $g < 0$ and the function $E(g)$ should have a cut 
along real negative semi-axis of $g$ plane. Summation with the help of Borel transformation results in a function 
having the left cut, besides the discontinuity on this cut is a smooth function of $g$.

    The Borel transformant $B(w)$ of function $E(g)$ is a series with coefficients $B_k = E_k/\Gamma (2k+b_0)$:
\begin{equation}
B(w) = \sum_{k=0}^{\infty } B_kw^k,
\end{equation}
where $b_0$ is an arbitrary constant. The choice of $b_0$ can affect, in principle, on the numerical results, but changing
of its value within interval $\sim [0.5\leq b_0 \leq 5]$ reveals weakly, so the choice of $b_0$ was made rather by convenience. The numerical
calculations in this work were performed at $b_0=3$.
The series (4) converges, as usual, within the circle $|w| < 1/a_n$. It is easy to check that the singularity of $B(w)$ is located at 
$w = -1/a_n$, substituting asymptotical coefficients $\tilde E_k$ in place of $E_k$. 
Energy of the level is related with the function $B(w)$ by an integral transform
\begin{equation}
E(g) = \int _{0}^{\infty }e^{-x}B(g x^2)x^{b_0-1}dx.
\end{equation}
For the numerical integration in the right hand side to be successful, an analytical continuation of $B(w)$ from its
convergence circle on the domain, containing the image of the entire real positive $w$ semi-axis is required.
For this aim we performed conformal mapping of the Borel variable $w$.
Many sufficiently effective versions of this mapping are appropriate. The main point is that the nearest singularity of 
the Borel transformant should be removed to infinity. Here we used the mapping 
\begin{equation}
y = \frac{a_nw}{1+a_nw}
\end{equation}
which was employed in the work~\cite{Suslov}. As is explained in~\cite{Suslov},
this transformation is optimal in the sense that it diminishes 
the influence of all possible singularities of $B(w)$ from the unphysical region. Transformation (6) is equivalent to 
the following series rearrangement
\begin{equation}
B(w) = \sum_{m=0}^{\infty }D_my^m ,
\ D_0 = B_0,\  D_m = \sum_{k=1}^{m}\frac{(m-1)!}{(k-1)!(m-k)!}\frac{B_k}{a^k}, \ m\geq 1.
\end{equation} 
To improve the convergence we applied Pad\' e summation to rearranged series (7)
\begin{equation}
B(w) \approx [M/N](y) \equiv P_M(y)/Q_N(y),
\end{equation}
where $P_M$ and $Q_N$ are polynomials of degree $M$ and $N$ respectively.
\par
We performed computations using various Pad\' e approximants and straightforward summation of the rearranged series (7).
To illustrate the influence of computational accuracy on summation results we compared ones made in double precision (16 decimal digits)
with these in quadruple precision (32 decimal digits).    

Some graphs of the obtained binding energy $\mathcal E(\mathcal H) =\displaystyle \frac{1}{2} \mathcal H - E(\mathcal H^2)$ as a function of parameter 
$\gamma \equiv n^3 \mathcal H$ are given in Figs. 1-3. As compared with the 
previous work~\cite{WGK},
the region of external field values for which these eigenvalues are successfully 
recovered is extended by a factor of about 5.
As usual the
precision of the sum considerably increases at lower $\mathcal H$ values. For instance in the case of $3d$
state with the help of approximants [N/N](y) in the range $27 \leq N \leq 32$ we get binding energies with 4 stable
decimal digits at $\gamma = 4$, with 6 digits at $\gamma = 3$ and with 12 digits at $\gamma = 1$.
Note that in the work~\cite{WGK} Pad\' e approximants were applied immediately to summation of divergent series (2).
These approximants imitate the discontinuity on the cut $g < 0$ by a set of 
delta-functions, and it is a very rough approximation. At the same time as a result of Borel summation the same discontinuity
is represented by a smooth function of $g$. Our calculations confirmed that mapping
(6) is indeed very effective:
after this mapping Pad\' e summation of the Borel transformant improves the 
convergence only a little and for some cases its straightforward summation appears to be sufficient -- see Figs. 1-3.

    One technical detail  is of principal importance for perturbation series summation by any method. The precision 
of the entire chain of computations must increase as the number of involved successive terms increases. 
This is simply a consequence of the fact that the sum, being of the order of unity, arises as a result of very large alternating sign terms compensation.

    It seems at first sight that the requirement of high precision is not necessary for the Borel transformant: all essential alternating 
sign coefficients $B_k$ have about the same order. But any numerical procedure of analytical continuation usually requires
high precision. Turning to series rearrangement (7) we see that binomial coefficients entering the sum for $D_k$ are
changing 20 orders of magnitude (in the present case). Obviously, we have enormous loss of precision performing the sum for $D_k$ in (7).
Therefore, if we want to use all $B_k$ up to 75th order,
the precision of $B_k$ coefficients should be better than
$\sim 10^{-20}$. In our calculations, the precision of $E_k$ and, consequently, the precision of $B_k$ was $\sim 10^{-30}$, so the precision of 
$D_k$ was decreasing from $10^{-30}$ at $k=0$ to about $10^{-10}$ at $k=75$.
\par
Let us turn now to the problem of restoring of the $E(g)$ dependence at large $g$ values. We shall focus on the ground state.
First of all we note that in work~\cite{Potekhin} an interpolation expression for the ground (tightly bound) state energy was obtained.
In spite of the multiple anticrossings
at $\mathcal H \leq 300$ and of the related computations complicating, the fit of~\cite{Potekhin} provides precision within $10^{-3}\div 10^{-2}$
in the range of $\mathcal H$ values $0.1\leq \mathcal H\leq 10^4$. 
\par
The ground level energy asymptotic at large $g$ (or, the same, at large $\mathcal H$) is given by
\begin{equation}
E(\mathcal H)\to \frac{1}{2}\mathcal H -\frac{1}{2}\ln^2\left(c\mathcal H\right) +...
\end{equation}
(see, for example,~\cite{Landau}). Here $c$ is a dimensionless constant.
First, we consider the possibility of restoring of the leading term parameters in (9) -- the power index and the
constant multiplier -- using perturbation theory. Methods applicable to this problem are considered in~\cite{Pop.Wein., Suslov}. Note that
for coming of the asymptotic into action the leading term in (9) should be large comparing with the correction term. One can look, for example, at
the results of work~\cite{Wang} (where the values of $\mathcal E(\mathcal H)$ were obtained by variational procedure) 
and ensure that only if $\mathcal H > 10^2$ then the binding energy will make less than 20\% of $\frac{1}{2}\mathcal H$.
So, we can speculate about  restoring of asymptotic parameters only if we succeeded in summation of $E(g)$ in this region of external fields.
But we failed to do this having used only 75 coefficients $E_k$, so the linear asymptotic couldn't be restored. This was confirmed in our attempts to
apply methods suggested in the works~\cite{Pop.Wein., Suslov}  -- no plausible result was obtained. 
In the method of Ref.~\cite{Suslov},
parameters of asymptotic of the function $E(g)$ were linked to behaviour of coefficients $D_k$ in dependence of their number $k$ at large $k$.
Namely, if $E(g)\to Cg^{\nu}$ at $g\to\infty$, then in our case we get similarly to~\cite{Suslov}
$$
D_k\to \frac{Ck^{\nu-1}}{a_n^k\Gamma(2\nu)\Gamma(2\nu +b_0)}.
$$ 
Then it was suggested to perform the fit of $C$ and $\nu$ using known $D_k$ and their errors by means of the $\chi^2$ method.
But in our case the value of $\chi^2$ in its
minimum was extremely large (about $10^{8}$ even if we tried to fit only 5 coefficients $D_k$ at statistical
error $\sigma=10^{-10}$, and we had no reason to increase this value of $\sigma$).
This result indicates that the asymptotic of $D_k$ comes into action at values of $k$ much larger than 75.
\par
The power index in $E(g)$ asymptotic could be traced also using the method of~\cite{Pop.Wein.}. This method concerns of taking of the limit of
the expression $\displaystyle \frac{wB^{\prime}(w)}{B(w)}$ at $w\to\infty$ (or, the same, the limit of 
$\displaystyle \frac{y(1-y)B^{\prime}(y)}{B(y)}$ at $y\to 1$), that gives exactly the value of $\nu$. But numerical calculation showed that in the region
where $B(y)$ was recovered (at $y$ close to $1$ we obviously should have increasing of error due to finite number of $D_k$ used)
we did not get reasonable precision for the limit value.
\par
Fig. 4 illustrates
precision of the linear asymptotic. The curve plotted represents the binding energy (we used here formula (6) of Ref.~\cite{Potekhin}) divided
by Landau energy versus external field.
One can see that at $\mathcal H \leq 100$ error of asymptotic is more than 20\%, and only at $\mathcal H\sim 1000$ precision reaches
level of 1\%.
So, it appears to be impossible to obtain asymptotic parameters corresponding to  Landau level for the Zeeman effect. 
\par
Now, the question arises, whether we can subtract Landau energy from $E(\mathcal H)$ and trace the second term of asymptotic.
But this term doesn't work even at $\mathcal H \sim 10^5$, which the graph of Fig. 5 is to explain. We plotted there the values of binding energy,
and the value of its logarithmic asymptotic.
We have chosen constant $c$ in such a way, that the value of logarithmic term $(1/2)\ln ^2(c\mathcal H)$ of the asymptotic 
coincides with the data of~\cite{Wang} at $\mathcal H = 10^5$.  This occurs at $c = 0.010$ and the smallness of this 
constant indicates that the value of $\mathcal H$ is too small for speaking about asymptotic.

And one can see that asymptotic curve and curve of exact
data have considerably different slopes. It means that we have to recover dependence of the energy at the values of external fields larger than
$10^5$. But it is really impossible having known only 75 coefficients $E_k$.
Thus, knowledge of 75 hypersucceptibilities did not allow to restore neither parameters of the Landau asymptotic nor 
logarithmic asymptotic of binding energy.

\begin{center}
\begin{figure}[h]
\includegraphics*[width=15cm, height=9cm]{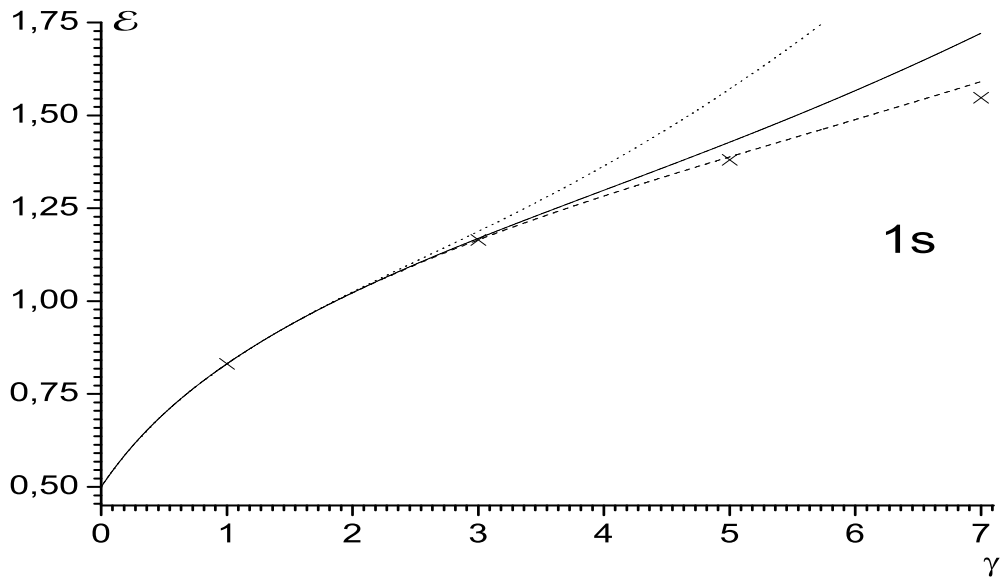}
\caption{Binding energy for 1s state in atomic units. The data evaluated in double precision (with help of Pad\'e approximant $[30/30]$
-- solid curve , by straightforward summation -- dotted curve) and in quadruple precision (with help of Pad\'e approximant $[30/30]$
-- dashed curve).
Crosses denote the data from Ref.~\cite{Wang})}
\end{figure}
\begin{figure}[h]
\includegraphics*[width=15cm, height=9cm]{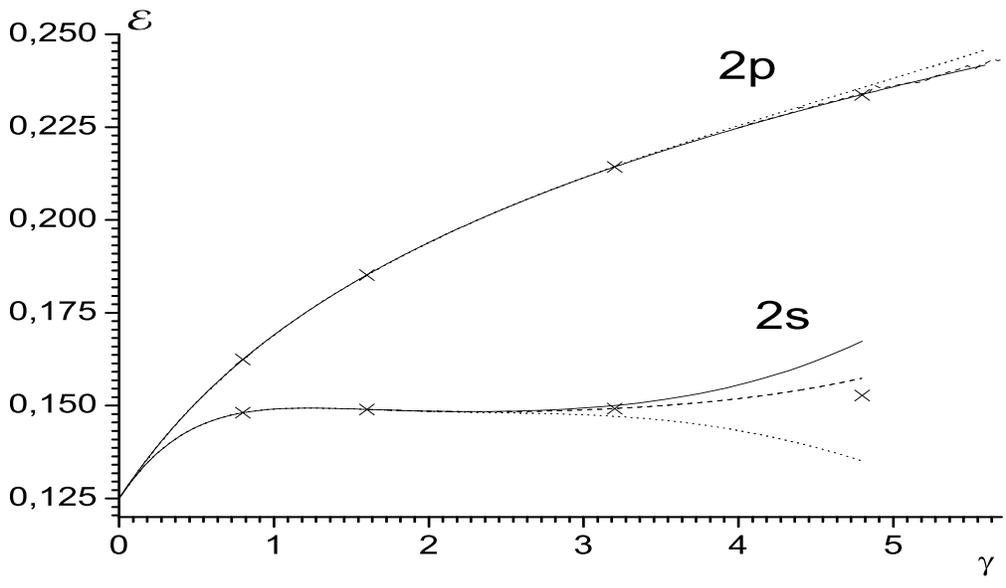}
\caption{Binding energy for 2s and 2p states in atomic units.
Notations are the same as in Fig. 1}
\end{figure}
\begin{figure}[h]
\includegraphics*[width=15cm, height=9cm]{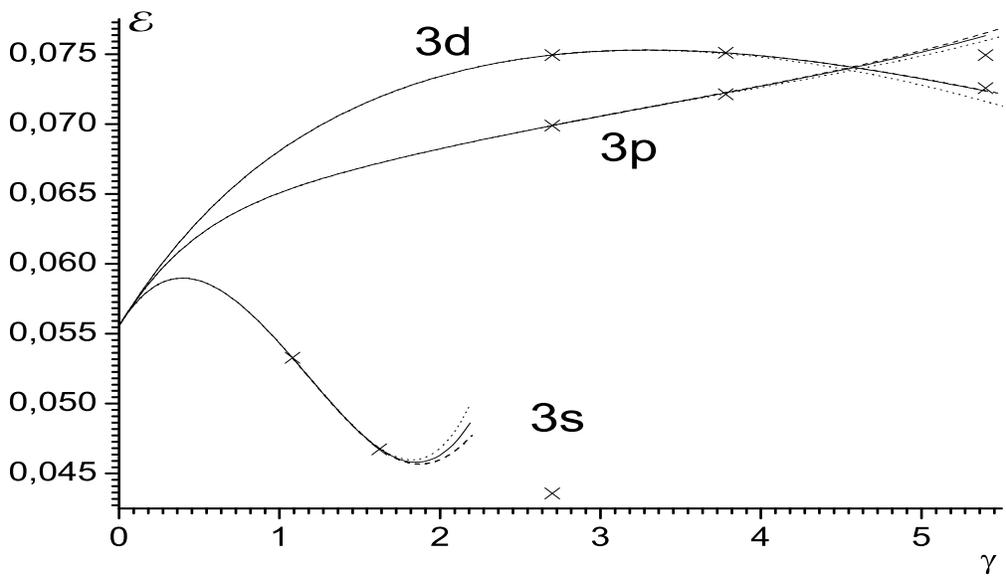}
\caption{Binding energy for 3s, 3p and 3d states in atomic units.
Notations are the same as in Fig. 1}
\end{figure}
\begin{figure}[h]
\includegraphics*[width=15cm, height=9cm]{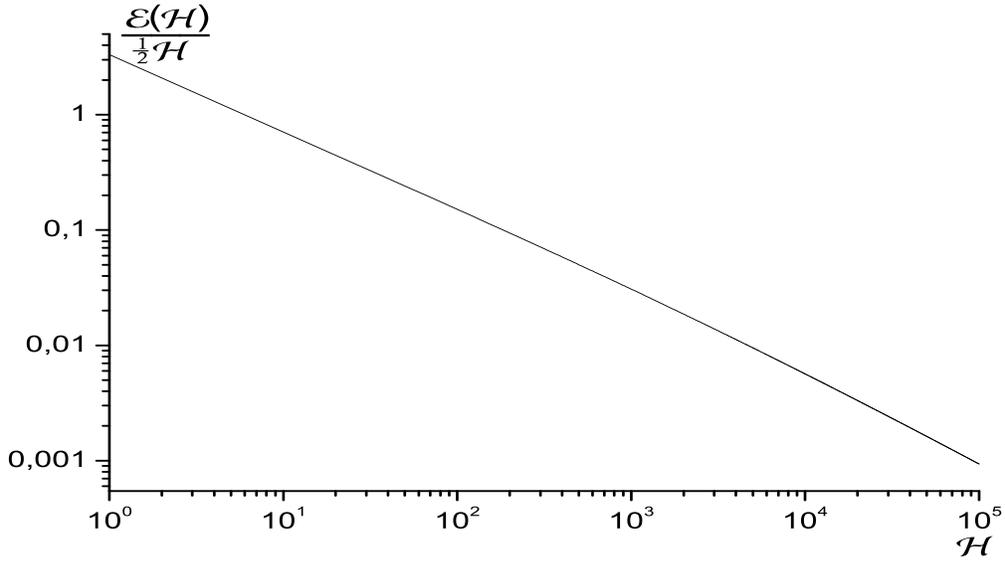}
\caption{Relative precision of the linear asymptotic}
\end{figure}
\begin{figure}[h]
\includegraphics*[width=15cm, height=9cm]{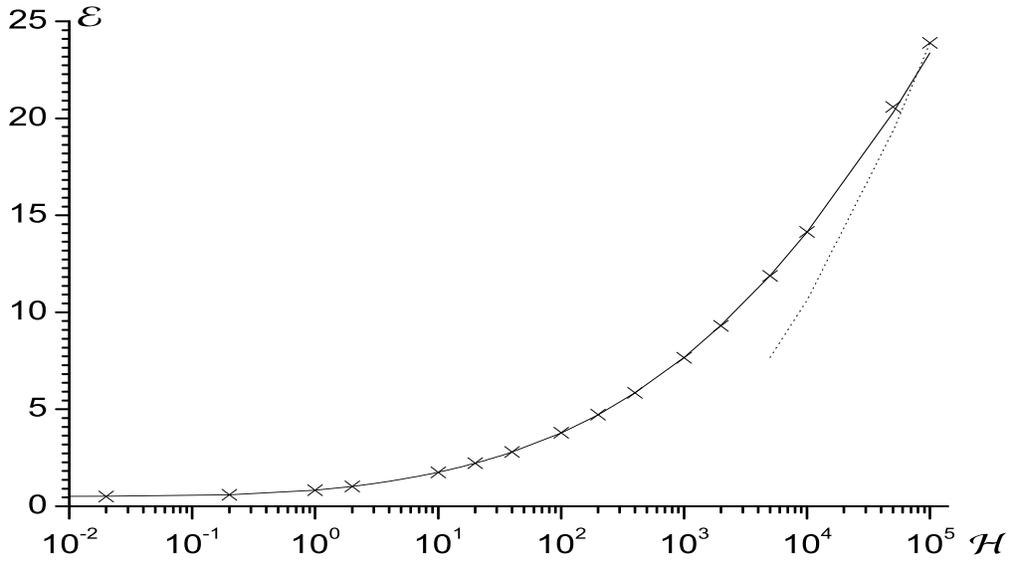}
\caption{Binding energy and its asymptotic. Solid curve is plotted using Eq. (6) from Ref.~\cite{Potekhin}. Crosses denote the data 
from Ref.~\cite{Wang}. Dotted curve is the logarithmic asymptotic with $c=0.010$}
\end{figure}
\end{center}

\end{document}